\documentclass[superscriptaddress,twocolumn]{revtex4}
\usepackage{amsmath,lscape,epsfig,ulem,color}

\def\beq{\begin{equation}}
\def\eeq{\end{equation}}
\def\beqa{\begin{eqnarray}}
\def\eeqa{\end{eqnarray}}
\def\ban{\begin{eqnarray*}}
\def\ean{\end{eqnarray*}}
\def\bi{\begin{itemize}}
\def\ei{\end{itemize}}

\begin{document}

\title{Light clusters and the pasta phase} 

\author{S. S. Avancini}
\affiliation{Depto de F\'{\i}sica - CFM - Universidade Federal de Santa
Catarina  Florian\'opolis - SC - CP. 476 - CEP 88.040 - 900 - Brazil}
\author{C. C. Barros Jr.}
\affiliation{Depto de F\'{\i}sica - CFM - Universidade Federal de Santa
Catarina  Florian\'opolis - SC - CP. 476 - CEP 88.040 - 900 - Brazil}
\author{L. Brito}
\affiliation{Centro de F\'{\i}sica Computacional, Department of Physics,
University of Coimbra,  P-3004 - 516   Coimbra, Portugal}
\author{S. Chiacchiera}
\affiliation{Centro de F\'{\i}sica Computacional, Department of Physics,
University of Coimbra,  P-3004 - 516   Coimbra, Portugal}
\author{D. P. Menezes}
\affiliation{Depto de F\'{\i}sica - CFM - Universidade Federal de Santa
Catarina  Florian\'opolis - SC - CP. 476 - CEP 88.040 - 900 - Brazil}
\author{C. Provid\^encia}
\affiliation{Centro de F\'{\i}sica Computacional, Department of Physics,
University of Coimbra,  P-3004 - 516   Coimbra, Portugal}

\begin{abstract}
The effects of including light clusters in nuclear matter at low densities
are investigated within four different parametrizations of relativistic models
at finite temperature. Both homogeneous and inhomogeneous matter 
(pasta phase) are described for neutral nuclear matter with fixed proton 
fractions. We discuss the effect of the density dependence of the symmetry
energy, the temperature and the proton fraction on the non-homogeneous matter
forming the inner crust of proto-neutron stars. It is shown that the number of
nucleons in the clusters, the cluster proton fraction and the sizes of the 
Wigner Seitz cell and of the cluster  are very sensitive to the  
density dependence of the symmetry energy.
\end{abstract}
\maketitle

\vspace{0.50cm}
PACS number(s): {21.65.+f, 24.10.Jv, 26.60.+c, 95.30.Tg}
\vspace{0.50cm}

\section{Introduction}

The formation of light clusters in  nuclear matter at 
low densities and its influence in the appearance and composition of pasta 
structures has been frequently discussed in the
literature \cite{ls91,shen,roepke,hor06,sumi08,typel09,hempel10,raduta10,ropke11,hempel11}.

Below saturation density homogeneous nuclear matter can become unstable against 
phase separation and several types of complex structures can be formed as a result 
of the competition between the strong and the electromagnetic interactions. This pasta phase
\cite{pethick,hashimoto,horo,watanabe,maruyama} 
{ is} found at densities of the order of 
0.001 - 0.1 fm$^{-3}$
\cite{pasta1}
in neutral nuclear matter, formed  by protons, neutrons and electrons, 
and  in a smaller density range 
in $\beta$-equilibrium stellar matter \cite{bao,pasta2}. At very low densities, up to 0.001 times the saturation 
density, and moderate temperatures, the few body correlations are still important and the system 
minimizes its energy by forming light nuclei like deuterons, tritons, helions and/or $\alpha$ particles. 
The appearance of these light clusters can modify the behavior of the
neutrinos in the expanding matter resulting from a supernovae core collapse
and  affect the cooling process of the protoneutron star. 

In a previous paper \cite{pasta-alpha} we have studied the influence of the
$\alpha$ particles both in homogeneous matter and in the
onset and structure of the pasta phase within the relativistic mean field 
approximation.
We have considered both free $\alpha$ particles and $\alpha$ particles interacting 
through an $\alpha-\omega$ meson coupling. This repulsive interaction is essential 
to avoid an overprediction of $\alpha$ particles above
$\rho\sim0.001$ fm$^{-3}$ and it is also the mechanism responsible for the 
dissolution of the $\alpha$ particle clusters in the approach considered.

Model dependences are more important for $ \rho>0.001$ fm$^{-3}$ when 
the $\alpha$ particle fraction differences between models may be as large as 
one order of magnitude or even larger. 
The effect of the temperature is to shift 
the maximum of 
the $\alpha$ particle distribution and the density of cluster dissolution to
larger densities. At the same time, the maximum values 
of the distributions themselves were shown to decrease with the increase 
of the temperature. The maximum 
values of the $\alpha$ particle distributions also decrease when the
proton fraction decreases. However, the proton fraction has no effect 
neither on the 
density localization of the maximum, nor on the density of 
dissolution of the clusters.

 It was also shown that in the pasta phase formed in asymmetric
nuclear matter the  $\alpha$ particle fraction increases with temperature.
 This is an interesting effect related to the proton fraction in the  
background gas  which increases with temperature for asymmetric matter.
It is important to test the above mentioned behaviors when other light
clusters are also included in the system.

In the present paper we extend our previous work \cite{pasta-alpha} by
considering also deuterons, tritons and helions. We study the distribution of
these light clusters as a function of the baryonic density for several
temperatures and proton fractions, and investigate the effect of the light 
clusters on the onset, type of structures and size of the heavy clusters 
($A > 4$) of the pasta phase. 
We  compute the mass and charge content of
the droplets formed in the droplet regime 
of the pasta phase and compare with other approaches. 

Our approach  to the description of the pasta phase is within what is known 
as the single nucleus
  approximation. The same approach was used in \cite{ls91,shen},
  however, only  alpha particles were included as light clusters.
 In \cite{hempel10} a statistical model consisting of an ensemble of nuclei
 and interacting nucleons in nuclear statistical equilibrium was proposed to
 describe supernova matter.  There, it was shown that
  the presence of light clusters besides alpha particles are of particular
  importance  at  low densities. A similar conclusion was drawn in
  \cite{raduta10} where  a phenomenological
  statistical model consisting of free nucleons described within a mean-field
  approximation and a loosely interacting cluster gas was formulated to describe
  matter in supernova explosions and proto-neutron stars. In all these works an
  excluding volume concept was used: in the first two works this was done with respect to
  the alpha particles and in the other two with respect to all nuclei or clusters.

The validity of the  two approximations, excluded volume and single nucleus
has already been discussed. Recently, 
the excluded volume approach was compared to two quantum many-body models in
\cite{hempel11} and it
  was shown that this approach is a bit crude at temperatures
  of the order of 5 MeV, although this occurs at densities where the
  composition of matter is dominated by heavy ions.
In \cite{burrows84} it was discussed that the EOS is not much affected by the single
  nucleus approximation, which describes matter composition in an average
  way, although, a correct distribution of nuclei may be important to describe
  correctly  the supernova-dynamics.

The approach we consider in the present work has the drawbacks of a
single nucleus approximation and does not include shell effects contrary to
statistical models \cite{hempel11}. Therefore, we will  restrict ourselves to temperatures above which shell
  effects are no longer important.
However, we would like to point out that  at densities close to the
  crust-core transition statistical models have difficulties in describing
  properly the medium effects on the nuclei and exotic structures such as the
  pasta phases contrary to the present approach. 

As referred before, we will avoid the excluded volume approximation and will
include  Pauli blocking 
and self-energy effects in a phenomenological
approach by including a meson-cluster interaction within a relativistic
mean-field formalism. A quantum statistical approach allows the calculation of
the medium
influence on the quasi energies of clusters
\cite{roepke,typel09,ropke11}. In particular, recently analytical fits to the
quasi-energy shifts of light nuclei were published  \cite{ropke11}. 
These results allow the determination of a better parametrization of the
cluster-meson interaction than the one used  in the present work. However, we
should point out that when the medium effects are more important the
contribution of light clusters is already very small. Moreover, we have shown
in \cite{pasta-alpha}
  that the fraction of alpha particles obtained with the prescription used
 in the present work  agrees with the results both of the virial equation
 proposed in \cite{hor06}  and the results of \cite{typel09} within a
 generalized mean-field model.

A relativistic mean-field approach is a phenomenological theory 
where the meson-nucleon and the
meson-meson interactions mimic the different contributions of a quantum
many-body formalism such as the Pauli blocking and  self-energies and it is
difficult to separate the different contributions which determine the
behaviour of the system. The same may be said with respect to the
meson-cluster interactions introduced in \cite{typel09,pasta-alpha}.
In a recent study \cite{marcio} the effect of the meson-cluster couplings  
in the
equation of state (EOS) of homogeneous nuclear matter  with clusters was
investigated within a zero temperature relativistic
mean-field approach. It was shown that the $\sigma$-cluster and
$\omega$-cluster couplings determine the behavior of the clusters, namely their
fraction and dissolution density.

In this work we use the relativistic mean field 
approximation and we consider four different parametrizations. 
We have chosen the NL3 \cite{nl3}, NL3$\omega \rho$ \cite{fsu}, FSU$_{\rm
  Gold}$ \cite{fsug} and IU-FSU \cite{iufsu} parametrizations of the
non-linear Walecka model (NLWM) \cite{sw}, which allow us to discuss the role of the density dependence
of the  symmetry energy on the properties of the  non-homogeneous nuclear EOS.

The paper is organized as follows: in section II we briefly review the 
formalism underlying the homogeneous neutral $npe$ matter with the
inclusion of the light clusters. In section  III the coexisting-phases method used to obtain the pasta phase is
briefly reviewed. In section IV our results are 
displayed and commented and in section V the final conclusions are drawn.

\section{The Formalism}

We consider a system of protons and neutrons with mass $M$
interacting with and through an isoscalar-scalar field $\phi$ with mass
$m_s$,  an isoscalar-vector field $V^{\mu}$ with mass $m_v$, an 
isovector-vector field  $\mathbf b^{\mu}$ with mass
$m_\rho$.  We also include 
{tritons} 
(${}^3$H, represented by $t$), 
{helions}
(${}^3$He, represented by $h$), $\alpha$ particles and deuterons ($d$).
A system of electrons 
with mass $m_e$ that makes matter neutral is also included.

The Lagrangian density reads:
$$
\mathcal{L}=\sum_{j=p,n,t,h}\mathcal{L}_{j}+\mathcal{L}_{{\alpha }}+
\mathcal{L}_d+ \mathcal{\,L}_{{\sigma }}+ \mathcal{L}_{{\omega }} + 
\mathcal{L}_{{\rho }}
$$
\begin{equation}
 + \mathcal{L}_{\omega \rho} + \mathcal{L}_e +\mathcal{L}_A,
\label{lag}
\end{equation}
where the Lagrangian density $\mathcal{L}_{j}$ is
\begin{equation}
\mathcal{L}_{j}=\bar{\psi}_{j}\left[ \gamma _{\mu }iD^{\mu }_j-M^{*}_j\right]
\psi _{j}  \label{lagnucl},
\end{equation}
the $\alpha$ particles and the deuterons are described as in 
\cite{typel09} with $\mathcal{L}_{{\alpha }}$ and $\mathcal{L}_d$ given, respectively, by
\begin{equation}
\mathcal{L}_{\alpha }=\frac{1}{2} (i D^{\mu}_{\alpha} \phi_{\alpha})^*
(i D_{\mu \alpha} \phi_{\alpha})-\frac{1}{2}\phi_{\alpha}^* M_{\alpha}^2
\phi_{\alpha},
\end{equation}
and
$$ \mathcal{L}_{d}=\frac{1}{4} (i D^{\mu}_{d} \phi^{\nu}_{d}-
i D^{\nu}_{d} \phi^{\mu}_{d})^*
(i D_{d\mu} \phi_{d\nu}-i D_{d\nu} \phi_{d\mu})$$
\begin{equation}
-\frac{1}{2}\phi^{\mu *}_{d} M_{d}^2
\phi_{d\mu},
\end{equation}
with 
\begin{eqnarray}
iD^{\mu }_j &=&i\partial ^{\mu }-g_{vj}V^{\mu }-\frac{g_{\rho j }}{2}
{\boldsymbol{\tau}}%
\cdot \mathbf{b}^{\mu } -{e\over 2}  \left( 1+\tau_3  \right) A^\mu , \label{Dmu} \\ 
& & j=p,n,t,h,\alpha,d  \nonumber \\
M^{*}_j &=&M -g_{s}\phi, \quad j=p,n \\
M^*_t&=&M_t=3M - B_t, \\
M^*_h&=&M_h=3M - B_h, \\
M^*_{\alpha}&=& M_{\alpha}= 4 M - B_{\alpha},\\
M^*_d&=& M_d= 2 M - B_d,
\label{Mstar}
\end{eqnarray}
with the binding 
energies given by $B_t=8.482$ MeV, $B_h=7.718$ MeV,
$B_{\alpha}=28.296$ MeV and $B_d=2.224$ MeV
and $g_{v j}= A_j g_v$ and $g_{\rho j}=|Z_j-N_j| g_{\rho}$, 
where $A_j$ is the mass number, $Z_j$, the proton number and 
$N_j$, the neutron number. 
Notice that in our model the cluster masses are fixed to constant values, they do not depend upon the temperature
and the density.
However, due to the cluster-meson interaction, we find (see Sec. IV) that the dissolution density of each cluster
increases with the temperature.

The electron Lagrangian density is given by
\begin{equation}
\mathcal{L}_e=\bar \psi_e\left[\gamma_\mu\left(i\partial^{\mu}+ e A^{\mu} 
\right)-m_e\right]\psi_e,
\label{lage}
\end{equation}
and the remaining terms in eq. (\ref{lag}) are 
\begin{eqnarray*}
\mathcal{L}_{{\sigma }} &=&+\frac{1}{2}\left( \partial _{\mu }\phi \partial %
^{\mu }\phi -m_{s}^{2}\phi ^{2}-\frac{1}{3}\kappa \phi ^{3}-\frac{1}{12}%
\lambda \phi ^{4}\right)  \\
\mathcal{L}_{{\omega }} &=&\frac{1}{2} \left(-\frac{1}{2} \Omega _{\mu \nu }
\Omega ^{\mu \nu }+ m_{v}^{2}V_{\mu }V^{\mu }
+\frac{1}{12}\xi g_{v}^{4}(V_{\mu}V^{\mu })^{2} 
\right) \\
\mathcal{L}_{{\rho }} &=&\frac{1}{2} \left(-\frac{1}{2}
\mathbf{B}_{\mu \nu }\cdot \mathbf{B}^{\mu
\nu }+ m_{\rho }^{2}\mathbf{b}_{\mu }\cdot \mathbf{b}^{\mu } \right)\\
\mathcal{L}_{\omega \rho } &=& \Lambda_v g_v^2 g_\rho^2 V_{\mu }V^{\mu }
\mathbf{b}_{\mu }\cdot \mathbf{b}^{\mu }\\
\mathcal{L}_{A} &=&-\frac{1}{4} F_{\mu \nu }F ^{\mu \nu }~,
\end{eqnarray*}
where $\Omega _{\mu \nu }=\partial _{\mu }V_{\nu }-\partial _{\nu }V_{\mu }$, 
$\mathbf{B}_{\mu \nu }=\partial _{\mu }\mathbf{b}_{\nu }-\partial _{\nu }
\mathbf{b}
_{\mu }-\Gamma_{\rho }(\mathbf{b}_{\mu }\times \mathbf{b}_{\nu })$ and $F_{\mu \nu }=\partial _{\mu }A_{\nu }-\partial _{\nu }A_{\mu }$.
The  parameters of the models are: the nucleon mass $M=939$ MeV,
the coupling parameters $g_s$, $g_v$, $g_{\rho}$ of the mesons to
the nucleons, the self interacting $\kappa$, $\lambda$ and $\xi$ 
constant couplings and the $\omega-\rho$ coupling $\Lambda_v$.
In the above Lagrangian density $\boldsymbol {\tau}$ is
the isospin operator. When the NL3 parametrization is used, $\Lambda_v$ is set
equal to zero. When the NL3$\omega \rho$ parametrization is chosen, we 
follow the prescription of \cite{fsu}, where the starting point was
the NL3 parametrization and the $g_\rho$ coupling was adjusted for each value
of the coupling $\Lambda_v$ in such a way that for
$k_F=1.15$ fm$^{-1}$ (not the saturation point) the symmetry energy is
25.68 MeV.
{ In particular, in this work we set $\Lambda_v$ to a moderately large value 
($\Lambda_v=0.03$, see Table \ref{tab_sets}). The comparison 
of NL3 and NL3$\omega\rho$ results is meant to show the effect of changing the isovector 
part of the Lagrangian only.}

\begin{figure*}[ht]
  \centering
\begin{tabular}{cc}
\includegraphics[width=0.35\linewidth]{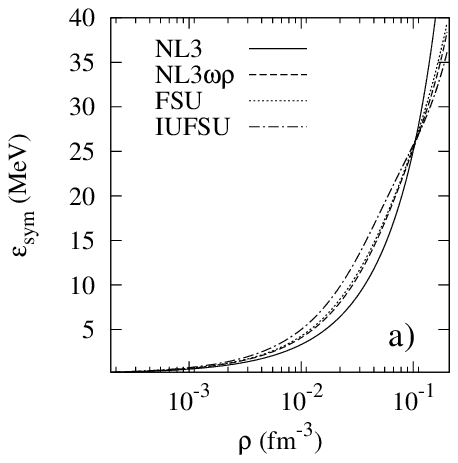} &
\includegraphics[width=0.35\linewidth]{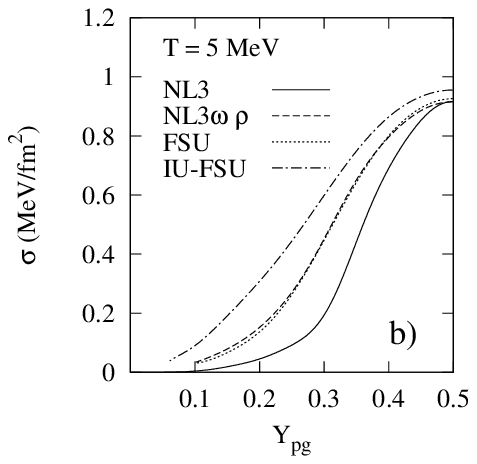}
\end{tabular}
\caption{Comparison of the a)  symmetry energy, b) surface tension at $T=5$
  MeV calculated with the models NL3, NL3$\omega\rho$, FSU$_{\rm
  Gold}$  and IUFSU. }
 \label{prop}
\end{figure*}

 The FSU$_{\rm Gold}$ \cite{fsug} parametrization was chosen because
it has the advantage of reproducing some of the results \cite{nosso}
obtained with more sophisticated density dependent hadronic models \cite{ddhd}
without the need of rearrangement terms. FSU$_{\rm Gold}$ combines the 
inclusion of the vector self-interaction term present in Ref.~\cite{tm1,ms} that is 
responsible
for explaining some observed properties at nuclear density and the isoscalar-
isovector coupling present in \cite{fsu}, capable of improving the density
dependence of the symmetry energy. An interesting comparison of the 
results for the neutron star mass-radius relation involving 
NL3 and FSU$_{\rm Gold}$ is given in 
\cite{piekarewics_2010}. We also consider the IU-FSU parametrization which 
keeps the main properties of FSU but was readjusted in order to allow 
for neutron star masses up to about 2.0 $M_{\odot}$ \cite{iufsu}. The 
parameter sets for the NL3, NL3$\omega\rho$, FSU$_{\rm Gold}$ and IU-FSU models
are shown in Table \ref{tab_sets}. Their corresponding
bulk nuclear matter properties are given in Table \ref{tab_bulk}.
In order to clarify the discussion,  we show in Fig. \ref{prop}, for the
models considered, the symmetry energy at densities below 
0.3 fm$^{-3}$ and the surface tension at
$T=5$ MeV as a function of the proton fraction. The surface tension was
  determined according to the parametrization given in eq. (\ref{sigpar}) and
  the Appendix. With the present choice of models we will be able to discuss
 the implications of the symmetry energy on the pasta phase. 
One expects, generally speaking, two types of effects:
a smaller $L$ corresponds to a larger surface tension for asymmetric matter
[see Fig.~\ref{prop}b) and Table \ref{tab_bulk}]; a larger ${\cal E}_{sym.}$ leads to a more 
isospin-symmetric liquid phase. We will see in section IV that 
in models with a larger surface tension the pasta phase sets in at higher 
densities and the drip of particles is unfavored, giving 
rise to a lower density background gas.

\begin{table}[h]
\caption{Parameter sets for the models used in this work. The masses of the mesons are in 
MeV and the other quantities are adimensional ($\kappa$ is given in nucleon mass units).}
\label{tab_sets}
\begin{center}
\begin{tabular}{lcccc}
\hline
        & NL3    & NL3$\omega \rho$ & FSU$_{\rm Gold}$ & IU-FSU \\
&   \cite{nl3} & \cite {fsu} & \cite{fsug} & \cite{iufsu}\\
\hline
\hline
$m_s$   & 508.194 &508.194 & 491.500 & 491.500 \\
$m_v$   & 782.501 & 782.501 & 782.500 & 782.500 \\
$m_\rho$ &763.000 & 763.000 & 763.000 & 763.000 \\
$g_s$   & 10.217  & 10.217 & 10.592 & 9.971 \\
$g_v$   & 12.868  & 12.868 & 14.302 & 13.032 \\
$g_\rho$ & 8.948  & 11.2766  & 11.767 & 13.590 \\
$\kappa$ & 4.384 & 4.384 & 1.7976 & 3.5695 \\ 
$\lambda$ & -173.31 & -173.31 & 299.11 & 2.926  \\
$\xi$   & 0.00 & 0.00 &0.06 & 0.03 \\
$\Lambda_{v}$ &0.00 & 0.03 & 0.03 & 0.046 \\
\hline
\end{tabular}
\end{center}
\end{table}

\begin{table}[h]
\caption{Nuclear matter properties at the saturation density
and zero temperature:
binding energy per nucleon $B/A$, density $\rho_0$, 
effective nucleon mass $M^*$ , incompressibility $K$,
symmetry energy ${\cal E}_{sym.}$, and slope $L$ 
of the symmetry energy ${\cal E}_{sym.}(\rho)$. 
}
\label{tab_bulk}
\begin{center}
\begin{tabular}{lccccccccc}
\hline
&  NL3 &  NL3$\omega \rho$ & FSU$_{\rm Gold}$ & IU-FSU \\

&   \cite{nl3} & \cite {fsu} & \cite{fsug} & \cite{iufsu}\\
\hline
\hline
$B/A$ (MeV) & 16.3  & 16.3  & 16.302 & 16.4\\
$\rho_0$ (fm$^{-3}$) & 0.148  & 0.148  & 0.148 & 0.155\\
$M^*/M$ & 0.60 & 0.60  & 0.62 & 0.62\\
$K$ (MeV) & 272 &  272 & 227.9 & 231.2\\
${\cal E}_{sym.}$ (MeV)  & 37.4  & 31.66 & 32.54 & 31.3 \\

$L$ (MeV) & 118.32 & 55.23  & 60.39 & 47.2 \\
\hline
\end{tabular}
\end{center}
\end{table}
After this discussion that motivates our choice 
of parametrizations, we go back to the description of the method.
From de Euler-Lagrange formalism we obtain coupled differential 
equations for the scalar, vector, isovector-scalar,
nucleon and cluster fields. In the static case there are no currents and the 
spatial vector components are zero. 
Moreover, we neglect, as usual, the Coulomb interaction in the case 
of homogeneous matter. In the calculation of the pasta
phase, its effect on the protons will be included.
In \cite{pasta1} a complete description of the 
mean-field and Thomas-Fermi approximations applied to different 
parametrizations of the NLWM is given and we do not repeat it here.
The equations of motion for the  fields are obtained and solved 
self-consistently and they can be read off \cite{pasta1,pasta2}.
The above mentioned equations of motion depend on the 
the  equilibrium  densities 
\begin{equation}
\rho=\rho_p + \rho_n + 4 \rho_{\alpha} + 2 \rho_d + 3 \rho_t + 3 \rho_h,
\label{rhotot}
\end{equation}

\begin{equation}
\rho_3=\rho_p - \rho_n - \rho_t +  \rho_h,
\end{equation}

\begin{equation}
\rho_s=\rho_{s_p} + \rho_{s_n}~.
\end{equation}
The quantities $\rho_{\alpha}$ and $\rho_d$ are discussed next, 
the fermionic densities are
\begin{equation}
\rho_i=\frac{1}{\pi^2} \int {p^2 dp}(f_{i+}-f_{i-}),\,\, i=p,n,t,h
\end{equation}
and the corresponding scalar densities are
\begin{equation}
\rho _{s_{i}}=\frac{1}{\pi^2} \int {p^2 dp}
\frac{M_{i}^{*}}{\epsilon^{\ast}_i}(f_{i+}+f_{i-})~.
\label{rhoscalar}
\end{equation}
The distribution functions are given by
\begin{equation}
f_{i \pm}=\frac{1}{1+\exp[(\epsilon^{\ast}_i({\mathbf p}) \mp \nu_i)/T]}\;,
\quad i=p,n,t,h
\label{distf}
\end{equation}
where
${\epsilon}^{\ast}_i=\sqrt{{\mathbf p}^2+{M^*_i}^2}$,
and the effective chemical potentials are
\begin{equation}
\nu_i=\mu_i - g_{vi} V_0 - \frac{g_{\rho i}}{2}~  \tau_{3 i}~ b_0 ,
\end{equation}
where 
\begin{equation}
\mu_{t}= \mu_p + 2 \mu_n, \quad \mu_h= 2 \mu_p + \mu_n,
\label{muf}
\end{equation}
and $\tau_{3 i}= \pm 1$ is the isospin projection for the protons (helions) and neutrons
(tritons) respectively.

In the present work the $\alpha$ particles and the deuterons are included as 
bosons and their chemical potentials are obtained from the proton and neutron 
chemical potentials
by imposing the chemical equilibrium,
as in \cite{typel09}: 
\begin{equation}
\mu_{\alpha}= 2 (\mu_p + \mu_n), \quad \mu_d=\mu_p + \mu_n~.
\label{mua}
\end{equation}
Their effective chemical potentials read
\begin{equation}
\nu_j=\mu_j - g_j V_0, \quad j=\alpha,d.
\end{equation}
The density of thermal $\alpha$ particles and deuterons are 
\begin{equation}
\rho_j=\frac{1}{\pi^2} \int {p^2 dp}(f_{j+}-f_{j-}),
\label{rhoalpha}
\end{equation}
with the boson distribution function given by
\begin{equation}
f_{j \pm}\,=\,\frac{1}{-1+\exp[(\epsilon_j \mp \nu_j)/T]},
\end{equation}
where $\epsilon_j=\sqrt{p^2+M_j^2}$.
We should point out that at low enough temperatures
 the $\alpha$s and deuterons contribute with two terms: a
  condensed fraction and a thermal contribution. 
Both contributions can be included
  explicitly in the present formalism. We only indicate the thermal
  contribution  because we have  limited our
  discussion to temperatures above Bose condensation of $\alpha$s or deuterons.

For the free electrons, their density and distribution functions
are the same as for the other fermions, where $\mu_e$ is the electron chemical 
potential and
$\epsilon_e=\sqrt{p^2+m_e^2}$. 
We always consider neutral matter and therefore the electron density is
equal to the total charge density of the charged particles.

In the description of the equation of state
of a system, the required quantities are the baryonic density $\rho$, energy
density $\cal E$, entropy density $\cal S$, 
pressure $P$ and free energy density $\cal F= {\cal E}-T {\cal S}$, 
and their expressions are explicitly given
in \cite{pasta1,pasta2}.

\section{Nuclear pasta}
In this section we 
describe briefly the coexisting-phases (CP) method to study the non-homogeneous
phase of
nuclear matter with a fixed proton fraction.
The basic idea is that matter can be organized into separated regions of higher and lower 
density, and the geometry of these regions is assumed to be very simple: 
a lattice of spherical droplets (bubbles), a plan of cylindrical rods (tubes) or an 
alternating sequence of slabs. 
The interface between regions is sharp, and it is taken into account by a surface term and
 a Coulomb one in the energy density. 
In the spirit of this approach, a single geometry will be the physical one  
for some given conditions (temperature, density and proton fraction).

As in \cite{pasta1,maruyama}, for a given total density $\rho$ and proton 
fraction, now defined including the protons present inside the
light clusters, the pasta structures are built with different 
geometrical forms in a background nucleon gas. This is achieved by calculating 
from the Gibbs conditions the density and the proton fraction of the pasta and
of the background gas, so that we have to solve simultaneously 
the following equations together with eqs. (\ref{muf}) and (\ref{mua}):
\begin{equation}
P^I=P^{II},\label{gibbs1}
\end{equation}
\begin{equation}
\mu_i^I=\mu_i^{II}, \quad i=p,n,t,h,\alpha,d, \label{gibbs2}
\end{equation}
\begin{equation}
f \rho_c^I +(1-f) \rho_c^{II}  = Y_{pg} \,\rho,\label{gibbs7}
\end{equation}
where I and II label the higher and the lower density phase respectively, 
and  $f$ is the volume fraction of 
phase I
\begin{equation}
f= \frac{\rho -\rho^{II}}{\rho^I-\rho^{II}} \label{f}~.
\end{equation}
The total baryonic density $\rho$ is given by eq. (\ref{rhotot}), $Y_{pg}$ is the global proton fraction given by
\begin{equation}
Y_{pg} = \frac{\rho_c }
{\rho}  
\end{equation}
and $\rho_c =\rho_p + 2 \rho_{\alpha} + \rho_d + \rho_t+ 2 \rho_h$ stands for the charge density. 

The density of electrons is uniform and taken as $\rho_e=Y_{pg}\, \rho$. 
The total pressure is given by  {the sum of the nuclear and electron partial pressures}
$P=P_{nucl}+P_e$,
where the nuclear contribution $P_{nucl}$ includes the kinetic contribution of
each type of particle (nucleons and light clusters) plus the meson contribution.
The total energy density of the system is given by
\begin{equation}
{\cal E}= f {\cal E}^I + (1-f) {\cal E}^{II} + {\cal E}_e +
{\cal E}_{surf} + {\cal E}_{Coul}~.
\label{totener}
\end{equation}
By minimizing the sum  ${\cal E}_{surf} + {\cal E}_{Coul}$ with respect
to the size of the droplet/bubble, rod/tube or slab one gets 
\cite{maruyama}
${\cal E}_{surf} = 2 {\cal E}_{Coul},$ and  
\begin{equation}
{\cal E}_{Coul}=\frac{2 F}{4^{2/3}}(e^2 \pi \Phi)^{1/3} 
\left[\sigma D (\rho_c^I-\rho_c^{II})\right]^{2/3},
\end{equation}
where $F$ is the volume fraction of the inner part 
($F=f$ for droplets, rods, slabs and $F=1-f$ for bubbles and tubes), 
$\sigma$ is the surface tension, 
$D$ is the dimension of the system, and 
$\Phi$ is a coefficient depending upon $F$ and $D$ \cite{pethick,maruyama}. 

Each structure is considered to be in the center of a charge neutral 
Wigner-Seitz  cell
constituted by neutrons, protons and electrons \cite{shen}. 
The Wigner-Seitz cell
is a sphere/cylinder/slab whose volume is the same as the unit BCC cell. 
In \cite{shen} the
internal structures are associated with heavy nuclei. Hence, the radius of the
droplet (rod, slab), $R_D$,  and of the Wigner-Seitz cell, $R_W$ are
respectively given by:
\begin{equation}
R_D=\left[ \frac{\sigma D}{4 \pi e^2 (\rho_c^I
- \rho_c^{II})^2 \Phi} \right]^{1/3},
\end{equation}
\begin{equation}
R_W=\frac{R_D}{F^{1/D}}. 
\end{equation}

Concerning the surface energy, the authors of \cite{watanabe2001} have 
shown how the surface energy affects the appearance of non-spherical pasta 
structures. Also the authors of \cite{maruyama} state that the appearance of 
the pasta phase essentially depends on the value of the surface tension. We 
have fixed the surface tension at different values
and confirmed their claim in \cite{pasta1,pasta2}, where the surface tension
 was parametrized in terms of the proton fraction
according to the functional proposed in  \cite{lattimer}, obtained
by fitting Thomas-Fermi and Hartree-Fock numerical values with a Skyrme force.
In \cite{pasta-alpha} a new prescription was used and the surface energy was
fitted to the results obtained from a relativistic Thomas-Fermi calculation and 
its dependence upon the temperature was taken into account.
A mathematical formula for $\sigma$ that gives accurate 
results for a broad  range of neutron excess and for temperatures up to 10 MeV
is
\begin{equation}
\sigma (x,T) = \tilde\sigma(x)\left[ 1-a(T)xT -b(T)T^2 -c(T)x^2 T\right]~,\label{sigpar} 
\end{equation}
where  $\tilde\sigma(x)$ is the surface tension at $T=0$ and $x=\delta^2$  stands for the squared relative neutron excess:
$$\delta = \frac{\rho_n-\rho_p}{\rho} = 1-2 Y_{pg}~.$$
In Table \ref{tabsig} the $\sigma$ parameters fitted to the 
Thomas-Fermi approximation results up to $T=10$ MeV are given. 
For $T=10$ MeV the parametrization is good if $Y_{pg}>0.2$. For
lower temperatures, the range of validity extends to lower $Y_{pg}$ 
values.

\begin{figure*}[ht]
  \centering
\begin{tabular}{cc}
\includegraphics[width=0.35\linewidth]{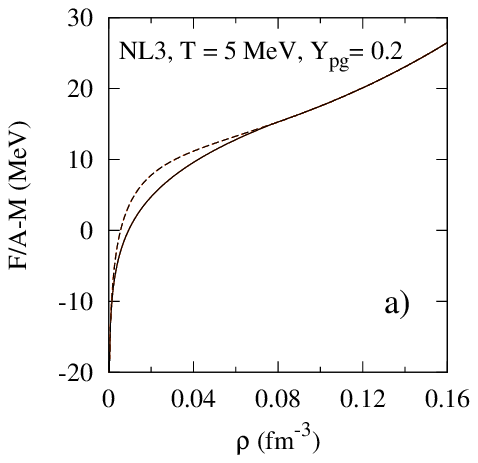} &
\includegraphics[width=0.35\linewidth]{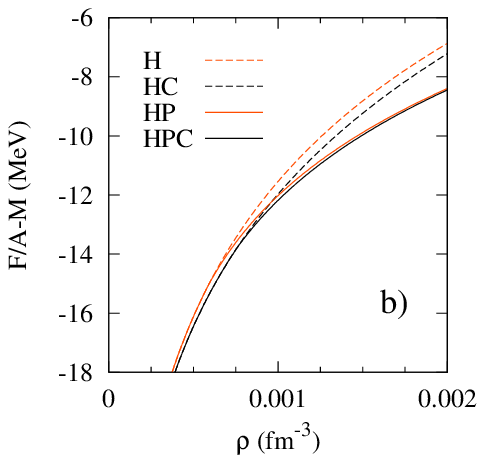}
\end{tabular}
\caption{(Color online) Free energy per particle as a function of the baryonic density 
for NL3 at $T=5$ MeV and $Y_{pg}=0.2$. Four constructions of the equation of state (H, HC, HP, HPC, see text) are shown
: a)
for subsaturation densities, the curves H and HC or HP and HPC are
indistinguishable;  b) at very low densities. }
 \label{free}
\end{figure*}

\section{Results and discussions}

In this section we present the results of this work.
In Fig. \ref {free} the free energy per particle is shown for  four different
constructions of the EOS:
homogeneous matter (H), homogeneous matter with clusters (HC), mixed 
homogeneous matter and pasta phase (HP)
and finally HP with clusters (HPC). 
Notice that with HP we mean that both non-homogeneous matter in five different
shapes (droplets, rods, slabs, tubes and bubbles) 
and homogeneous matter are computed and the phase with  the lowest free energy 
per particle is the physical one.

The free energy is clearly lowered by both the inclusion of pasta and clusters.
However, the pasta phase affects the results in a large density range and 
leads to a decrease of a few MeV [Fig. 2a)], whereas 
the clusters are visible only at very low density and lead to a decrease of 
$\lesssim$ 1 MeV [Fig. 2b)]. Their effect  is even smaller in the
pasta phase range. It is also seen, 
as already discussed in \cite{typel09}, that the pasta phase sets on at a larger density when light
    clusters are included. 

The inclusion of the light clusters in the EOS improves the CP method. In fact 
the way the surface energy enters in the CP calculation doesn't allow the 
appearance of these light nuclei but, as we have seen, their presence is important 
just below the onset of the pasta phase.  We will later verify whether the
inclusion of the light clusters explicitly does not correspond to a double
description of the small clusters in the pasta phase. 
In fact, if we include light clusters and obtain droplets formed by four or less nucleons
the same physical object is being described simultaneously in two
distinct ways.

\begin{figure*}[ht]
  \centering
\begin{tabular}{c}
\includegraphics[width=0.7\linewidth]{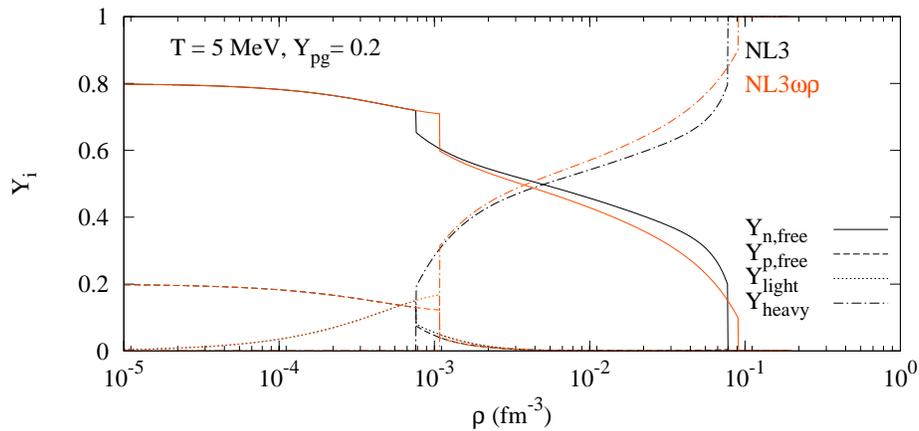}
\end{tabular}
\caption{(Color online) Fractions of nucleons as a function of the density,
for the NL3 and NL3$\omega \rho$ parametrizations 
at $T=5$ MeV and $Y_{pg}=0.2$.}
\label{ynl3wr}
\end{figure*}

\begin{figure*}[ht]
  \centering
\begin{tabular}{c}
\includegraphics[width=0.7\linewidth]{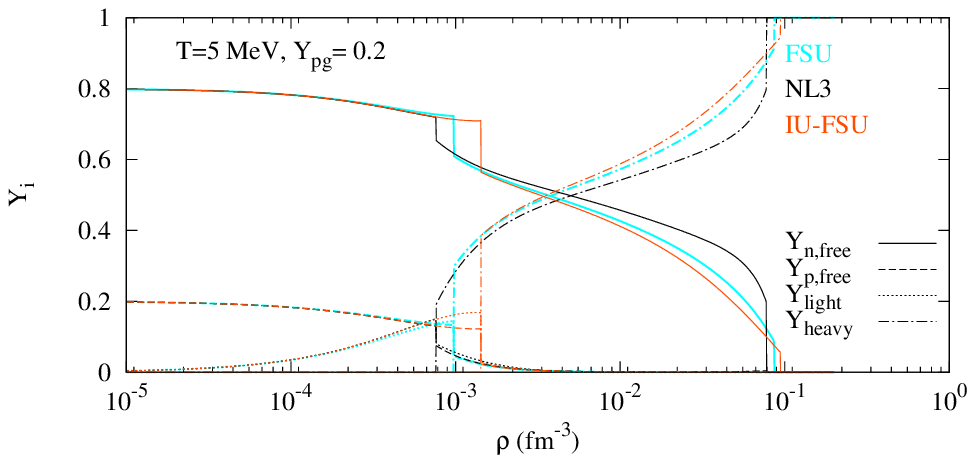}
\end{tabular}
\caption{(Color online) Fractions of nucleons as a function of the density, for the NL3, FSU$_{\rm
  Gold}$ 
  and IU-FSU parametrizations at $T=5$ MeV and $Y_{pg}=0.2$.}
 \label{yseveral}
\end{figure*}

In the more complete case (i.e., HPC), we can see how 
the fraction of nucleons belonging to different structures changes with the
total baryonic density. Nucleons can be free (not clusterized), belong
to a light cluster ($A\le 4$) or to a heavy one. In our formalism, the role of heavy 
clusters is played by the high density part of pasta structures.
In the regimes of temperature and proton fraction where a non-homogeneous phase
 appears, the following sequence is generally found: homogeneous matter at very low 
baryonic density, pasta phase at intermediate densities and then again 
homogeneous matter at higher densities.

In the non-homogeneous matter range, the fraction of a generic constituent has a contribution
from the high density (phase I) and one from the low density phase (phase II):
\begin{equation}
Y_i=Y_{i}^If\frac{\rho^I}{\rho}+Y_{i}^{II}(1-f)\frac{\rho^{II}}{\rho}
\end{equation}
where $i=p,n,\alpha,h,t,d$ and 
$f$ is the volume fraction of the denser phase defined in (\ref{f}).

Therefore, we classify the fraction of nucleons as follows:
\begin{equation}
\left\{
\begin{array}{lll}
Y_{p,free} & = & Y_p\\
Y_{n,free} & = & Y_n\\
Y_{light} & = & Y_\alpha+Y_h+Y_t+Y_d\\
Y_{heavy} & = & 0
\end{array}
\right.
\end{equation}
in homogeneous matter at low density,
\begin{equation}
\left\{
\begin{array}{lll}
Y_{p,free} & = & Y_{p}^{II}(1-f)\frac{\rho^{II}}{\rho}\\
Y_{n,free} & = & Y_{n}^{II}(1-f)\frac{\rho^{II}}{\rho}\\
Y_{light} & = & (Y_{\alpha}^{II}+Y_{h}^{II}+Y_{t}^{II}+Y_{d}^{II})(1-f)\frac{\rho^{II}}{\rho}\\
Y_{heavy} & = &
f\frac{\rho^{I}}{\rho}
\end{array}
\right.
\end{equation}
in the non-homogeneous phase,
and finally
\begin{equation}
\left\{
\begin{array}{lll}
Y_{p,free} & = & 0\\
Y_{n,free} & = & 0\\
Y_{light} & = & 0\\
Y_{heavy} & = & 1
\end{array}
\right.
\end{equation}
in the high density homogeneous phase.

In Fig. \ref {ynl3wr}  the fractions of nucleons are shown for the NL3 and
NL3$\omega \rho$ parametrizations at $T=5\,$MeV, and $Y_{pg}=0.2$. 
For both parametrizations, matter in the low density regime is formed by not-clusterized neutrons and protons and a light-cluster fraction. 
The magnitude of this fraction depends on the temperature. At $\rho\sim  0.001$ fm$^{-3}$ the pasta phase sets on. 
The background low density gas is constituted mainly by neutrons with a small fraction of protons and light clusters. 
For densities above $0.08$ fm$^{-3}$ there is a transition to dense homogeneous matter, interpreted in the present approach as an infinite cluster.  

\begin{figure*}[tbh]
  \centering
\begin{tabular}{c}
\includegraphics[width=0.7\linewidth]{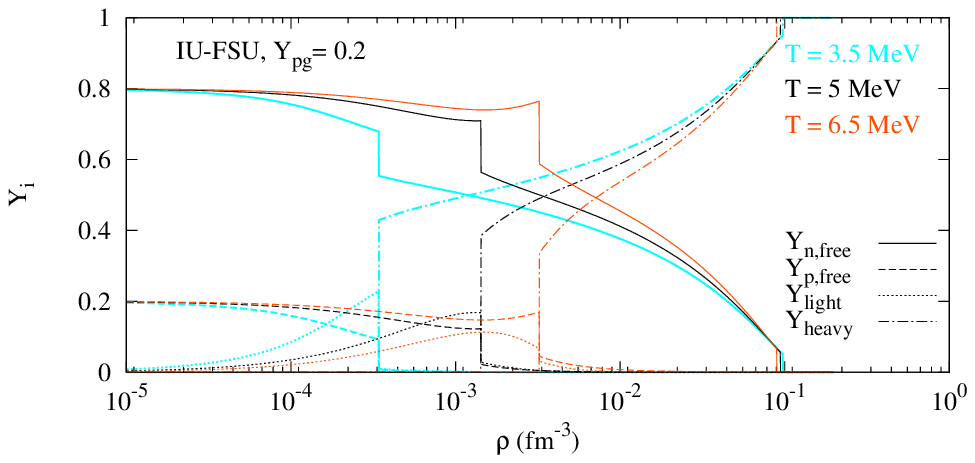}
\end{tabular}
\caption{(Color online) Fractions of nucleons as a function of the density, for the IU-FSU
  parametrization at $Y_{pg}$=0.2 and three different temperatures.}
 \label{ytemp}
\end{figure*}

\begin{figure*}[ht]
  \centering
\begin{tabular}{c}
\includegraphics[width=0.7\linewidth]{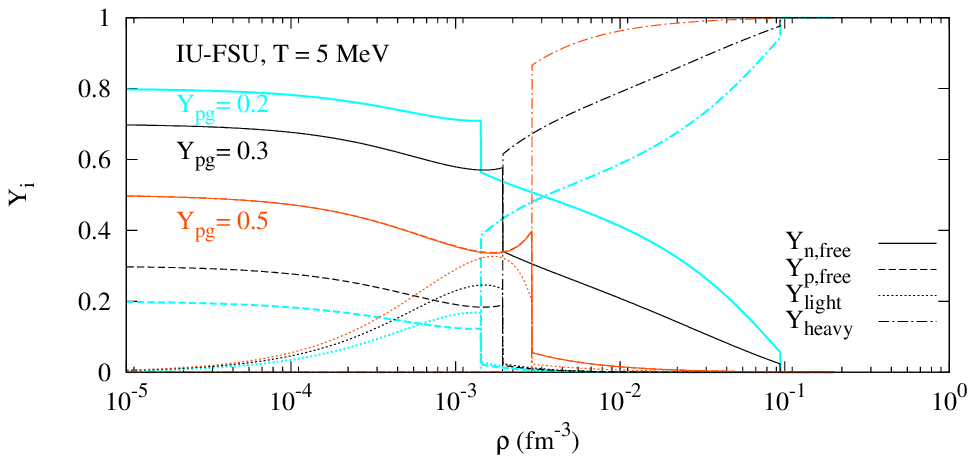}
\end{tabular}
\caption{(Color online) Fractions of nucleons as a function of the density, for the IU-FSU parametrizations at $T=5\,$MeV and several proton fractions.}
\label{fracyp}
\end{figure*}

The differences between the results obtained for NL3 and NL3$\omega\rho$ 
can be interpreted as a consequence of the effect of the density
dependence of the symmetry energy. Both NL3 and NL3$\omega\rho$ have the same
isoscalar properties. However they differ on the isovector channel, namely the
symmetry energy and its slope at saturation are different, NL3$\omega\rho$
having a smaller symmetry energy and smaller slope at
saturation.  Below $\rho=0.1$ fm$^{-3}$   the symmetry energy is larger for NL3$\omega\rho$ while
the opposite occurs above that density [see Fig. \ref{prop}b)].
It is seen that both models behave in the same way below the onset of the
pasta phase:  the proton, neutron and light cluster fractions are
practically equal. However there are noticeable differences on the pasta
phase.  The most important ones are the onset density of the pasta phase and
the fraction of nucleons in the cluster. A smaller symmetry energy slope
shifts the onset of the pasta phase of NL3$\omega\rho$ to larger densities because it
gives rise to a larger surface energy that hinders the formation of pasta
structures. A larger surface energy also gives rise to a smaller fraction of
neutrons outside the cluster because it is more difficult for neutrons to
drip out. Since neutrons have an important role on the cooling of the
crust, the fraction of free neutrons on the pasta phase range will
certainly affect the cooling and transport properties of the crust.

\begin{figure}[ht]
  \centering
\begin{tabular}{c}
\includegraphics[width=0.9\linewidth]{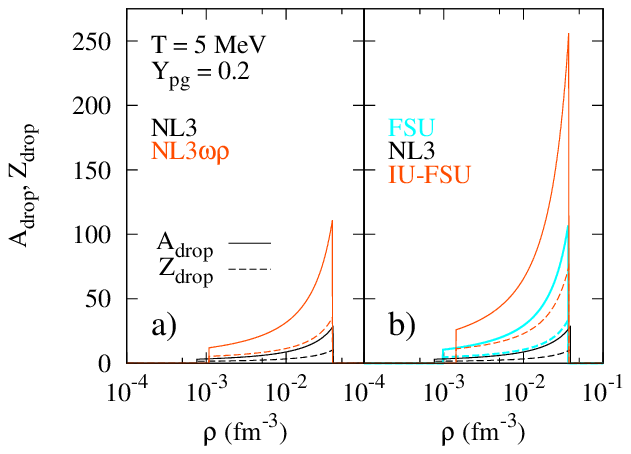}
\end{tabular}
\caption{(Color online) $A$ and $Z$ in a droplet at $T=5\,$MeV and $Y_{pg}=0.2$: a) NL3 and NL3$\omega\rho$ parametrizations; b) NL3, FSU$_{\rm
  Gold}$  and IU-FSU parametrizations.}
 \label{Zdroplet}
\end{figure}

\begin{figure}[ht]
  \centering
\begin{tabular}{c}
\includegraphics[width=0.9\linewidth]{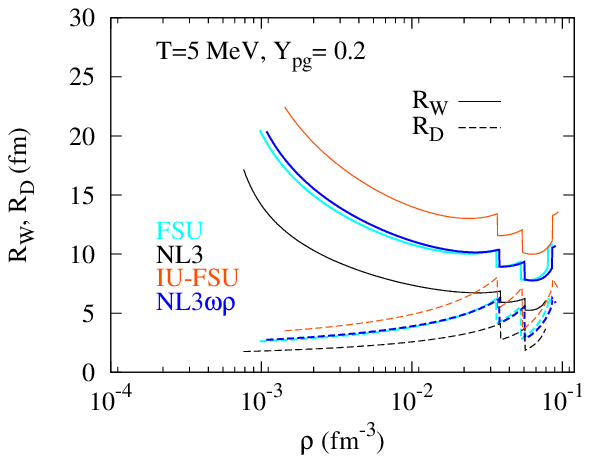}
\end{tabular}
\caption{(Color online) Droplet and Wigner Seitz cell radius of the spherical clusters for NL3, NL3$\omega\rho$,
  FSU$_{\rm
  Gold}$  and IU-FSU calculated at $T=5$ MeV and $Y_{pg}=0.2$.}
\label{radius}
\end{figure}

We plot in Fig. \ref {yseveral} the fractions of nucleons predicted by the FSU$_{\rm
  Gold}$ 
and IU-FSU parametrizations compared to the NL3 results. There are some
similarities between these results and those obtained with NL3$\omega \rho$.
The onset of the
pasta phase density occurs for the IU-FSU parametrization at values slightly
higher than in FSU$_{\rm
  Gold}$  and much higher than in NL3, due to its smaller $L$  and, therefore, 
larger surface tension.
 At low densities, the main
differences among the three models occur at $\rho\sim 0.001$ fm$^{-3}$ with
different proton and light cluster fractions. These differences are
  mainly due to  differences in the density dependence of the isoscalar channel of the EOS.
The FSU$_{\rm
  Gold}$  parametrization has the largest $g_s$ coupling, so, since nucleons have a
  smaller effective mass, and a larger binding energy per particle, the formation of light clusters
  is not so favored as in a model where the nucleon effective mass is larger. 
However, a small fraction of light clusters is still observed in FSU$_{\rm
  Gold}$ , because the onset
 of pasta phase occurs at densities larger than in NL3.

The effect of the temperature on the nucleon ratios is examined 
in Fig. \ref{ytemp} where the results for $T=3.5,\, 5,$ and 6.5 MeV are shown for the
  IU-FSU parametrization with $Y_{pg}=0.2$. The main features of increasing the temperature 
can be summarized in three points: a) the onset of the pasta phase (of the core)
 is shifted to higher (smaller) densities, as already discussed in \cite{silvia10};
b) the low density gas of the pasta phase has a larger number of particles; c)
the light clusters contribution is smaller at densities below the pasta phase
onset and larger on the low density background gas of the pasta phase. These
behaviors all occur because when the temperature increases the instability
region decreases and there is a larger number of nucleons that drip out of the
dense clusters.

\begin{figure}[ht]
  \centering
\begin{tabular}{c}
\includegraphics[width=0.9\linewidth]{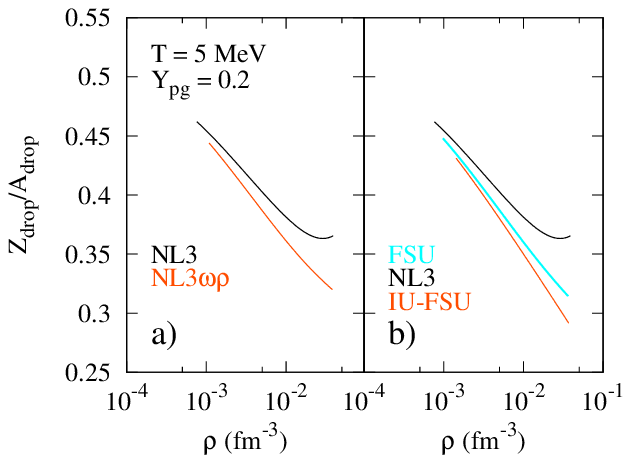}
\end{tabular}
\caption{(Color online) The ratio $Z/A$ in a droplet at $T=5\,$MeV and $Y_{pg}=0.2$: a) NL3 and NL3$\omega\rho$ parametrizations; b) NL3, FSU$_{\rm
  Gold}$  and IU-FSU parametrizations.}
 \label{ratio1}
\end{figure}

The effect of the global proton fraction on the pasta phase is clearly seen in
Fig. \ref {fracyp}, where the different constituent fractions  at densities below 0.1 fm$^{-3}$ are
plotted for $Y_{pg}=0.2, 0.3,$ and 0.5 for the IU-FSU parametrization at $T=5$ MeV. The onset of the pasta phase occurs at larger densities
for the more symmetric matter. This is possibly a trend  due to
the method used for the pasta phase clusters with a zero thickness surface. In fact, in the
pasta calculation within the Thomas Fermi approach, where the surface is
described in a self-consistent way, the opposite occurs \cite{silvia10}.
 However, if we also consider the light clusters, the onset of the
 clusterization starts at lower densities for the larger proton fractions. 
In the pasta phase, the larger proton fractions give rise to  clusters with a
 larger number of nucleons immersed in a  gas of nucleons and
 light clusters with a lower density. 

\begin{figure}[ht]
  \centering
\begin{tabular}{c}
\includegraphics[width=0.9\linewidth]{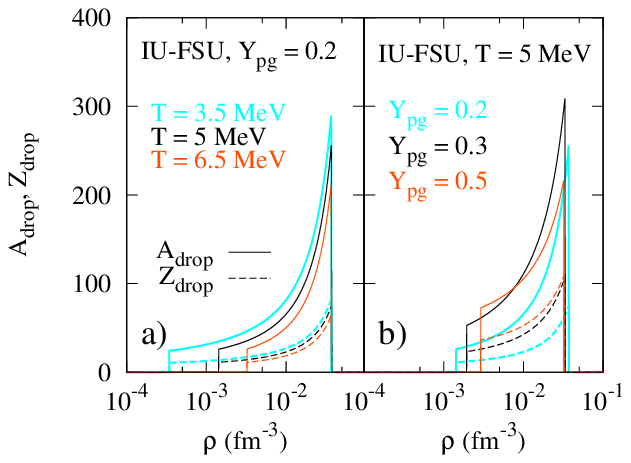}
\end{tabular}
\caption{(Color online) $A$ and $Z$ in a droplet for IU-FSU: a) $Y_{pg}=0.2$ and $T=$3.5, 5 and 6.5 MeV;
 b) $T=5\,$MeV and $Y_{pg}=$0.2, 0.3 and 0.5.}
 \label{ZdropletTY}
\end{figure}

\begin{figure}[ht]
  \centering
\begin{tabular}{c}
\includegraphics[width=0.9\linewidth]{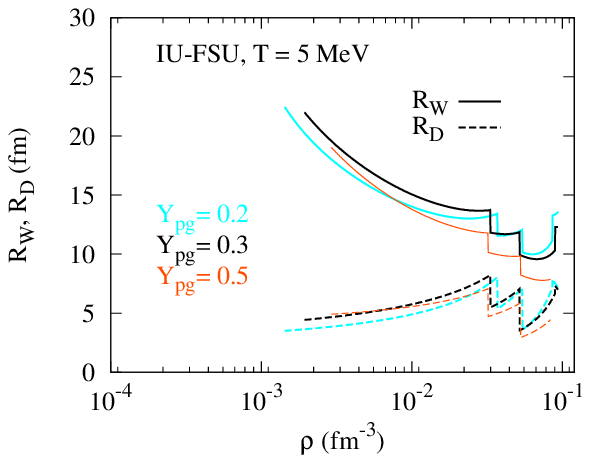}
\end{tabular}
\caption{(Color online) Droplet and Wigner Seitz cells radii of the spherical clusters for IU-FSU
  calculated at $T=5$ MeV and $Y_{pg}=0.2,\,0.3$ and 0.5.}
\label{radius2}
\end{figure}

These results allow us to make some comments on the possible consequences
 of the evolution of a protoneutron star. After the supernova explosion 
neutrinos
  are trapped inside the star and we can consider that the lepton fraction is
  approximately constant, taking a maximum value of 0.4 and decreasing as
  neutrinos leave the star. A lepton fraction of 0.4 corresponds to a proton
  fraction of the order of 0.3. During deleptonization the proton fraction of
  stellar matter decreases and after total deleptonization at densities below
  0.08 fm$^{-3}$ the proton fraction is below 0.1.
From the behavior obtained in Fig. \ref {fracyp} we conclude that 
 during deleptonization the number of nucleons in the  clusters decreases, 
the low density background gas of nucleons and light clusters increases and 
the onset of the clusterized phase shifts to larger densities.

\begin{figure}[ht]
  \centering
\begin{tabular}{c}
\includegraphics[width=0.9\linewidth]{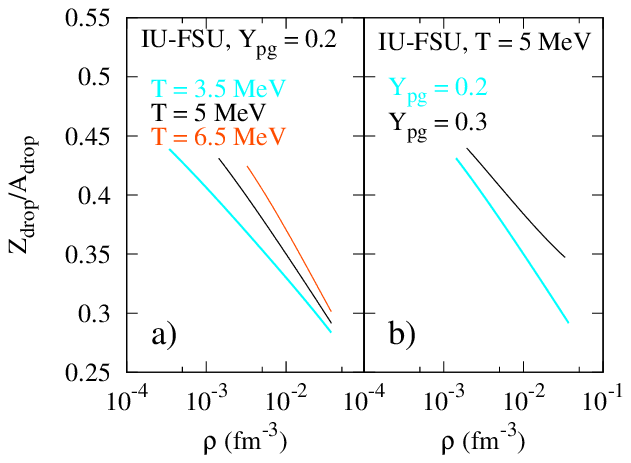}
\end{tabular}
\caption{(Color online) The ratio $Z/A$ in a droplet for IU-FSU: a) $Y_{pg}=0.2$ and $T=$3.5, 5 and 6.5 MeV;
 b) $T=5\,$MeV and $Y_{pg}=$0.2, 0.3 and 0.5.}
 \label{ratio2}
\end{figure}

Next we analyze the properties of the clusters formed in the pasta phase. We
consider the droplet geometry because of the finite size of this
structure. In particular we discuss the effects of temperature, isospin
asymmetry and the density dependence of the symmetry energy  on the
number of nucleons and protons in the droplets, on the onset density
of this geometry and the transition density to the rod geometry.

We denote by  $A_{\rm drop}$  the number
of nucleons belonging to a droplet (the type of structure that appears at the
lowest densities in the non-homogeneous phase). 
We first  check if we are including correctly the light clusters: 
in fact, we have to check that 
$A_{\rm drop}>4$, to confirm that classifying these droplets as heavy clusters is correct. 
Secondly, we can compare our results to those obtained in other
approaches, \cite{hempel10,raduta10}. 
We calculate this quantity as 
\begin{equation}
A_{\rm drop}=\frac{4 \pi}{3}\,R_D^3\,[\rho^I- \rho^{II}(Y_p^{II}+Y_n^{II})]\,.
\end{equation}
In a similar way, we can compute $Z_{\rm drop}$, the charge content of a
droplet. 
Next we show several figures including our results for both 
$A_{\rm drop}$ and  $Z_{\rm drop}$ for various models, 
temperatures and proton fractions.

In Fig. \ref{Zdroplet} we display the results for different parametrizations:  NL3
and NL3$\omega \rho$ in Fig. \ref{Zdroplet}a), NL3, FSU$_{\rm
  Gold}$  and IU-FSU in Fig. \ref{Zdroplet}b).
Here and in the following analogous figures, the solid lines represent $A_{\rm drop}$, 
the dashed ones
$Z_{\rm drop}$.
The onset of the droplet
phase is characterized by a discontinuity on the number of nucleons inside the
cluster: it is necessary a minimum number of nucleons to compensate the
surface energy, that is larger for NL3$\omega \rho$. In a Thomas-Fermi calculation, where the surface energy is
calculated self-consistently we may expect a less discontinuous
behavior. A change in  the isovector
channel of the model as in NL3$\omega\rho$ leads to a large effect on the
number of nucleons, increasing this number to more than the
double. 
 As discussed before, a smaller symmetry energy slope corresponds to a larger
  surface energy and neutrons do not drip out so easily. The
number of nucleons obtained within NL3$\omega\rho$  is consistent with the 
results of \cite{hempel10} within a statistical model, based on the TMA 
parametrization \cite{tma}.  We should point out that for NL3 the 
separation that has been done in light clusters and heavy clusters breaks 
down for a small proton fraction because, at low densities, the size of the heavy clusters equals the 
size of the light clusters.  
The models which include an isoscalar-isovector coupling $\Lambda_v$ present
larger  nuclei, the heaviest ones corresponding to IU-FSU. In these models, the appearance of
the heavy clusters occurs  at similar densities, larger than that obtained for NL3.
 This again is due to the fact that
NL3 has the largest symmetry energy slope at these densities.

In Fig. \ref{radius}
we plot the radius of the Wigner Seitz cell together with  the droplet
radius as a function of density for NL3, NL3$\omega\rho$, FSU$_{\rm
  Gold}$  and IU-FSU.
The ordering of the radii obtained in the different parametrizations
reflects perfectly the ordering of their surface tensions [cfr. Fig. \ref{prop} b)], that, 
in turn, is closely linked to the symmetry energy density dependence [cfr. Fig. \ref{prop} a) and Table \ref{tab_bulk}].
NL3 has by far the smallest surface energy at $Y_{pg}=0.2$, while IU-FSU
has the largest, see Fig. \ref{prop}b): correspondingly, NL3 has the smallest  
Wigner Seitz cell and droplets and IU-FSU the largest ones, as shown in Fig. \ref{radius}.

In Fig. \ref{ratio1} the ratio $Z_{\rm drop}/A_{\rm drop}$
  is plotted at a function of density for the different models we are
  comparing. We conclude that this ratio decreases with density and is model
  dependent. A decrease of the proton fraction of the clusters with density
  was also obtained in \cite{raduta10}. 

The models with a smaller symmetry
  energy slope have smaller proton fractions. A smaller slope implies that
  neutrons drip out of the cluster with more difficulty giving rise to
 neutron richer clusters.  
 Also,  a smaller slope results in a smaller
${\cal E}_{sym.}(\rho)$ above $0.7\rho_0$ 
[$\simeq 0.1$ fm$^{-3}$, where approximately all the curves cross, see Fig. \ref{prop} a)].
Since the density inside the droplets is
  between $0.7\rho_0$ and $\rho_0$,   a smaller symmetry energy
  favors  less symmetric clusters.

The dependence of $A_{\rm drop}$ and $Z_{\rm drop}$ on the temperature for $Y_{pg}=0.2$, 
and on the proton fraction, for $T=5\,$MeV is plotted in
Fig. \ref {ZdropletTY} a) and b), respectively, 
for the IU-FSU parametrization.
Decreasing the temperature
increases slightly the number of nucleons in the
clusters and strongly decreases the onset density. The transition to the rod 
geometry seems to be temperature independent.

Isospin asymmetry  affects  the number of nucleons in the
cluster in a non linear way as is seen in the right panel of Fig. \ref{ZdropletTY} for the IU-FSU
parametrization. At the onset of the pasta phase the size of the clusters
is smaller for the smaller proton fractions. However, there is a faster
increase of the cluster size with density when the proton fraction is smaller. In
particular, at the transition to the rod geometry we get  smaller droplets in
symmetric matter. This behavior is also obtained for FSU$_{\rm Gold}$  and
NL3$\omega\rho$. However, for NL3, we get a systematic behavior, the smaller the proton
fraction the smaller the number of nucleons in the cluster. 
 In \cite{raduta10}, using the Skyrme interaction SKM$^*$ with a quite low
 symmetry energy slope at saturation ($L$=45 MeV),  a cluster  size
 independent  of
the proton fraction was obtained. 
On the other hand, the authors of  \cite{hempel10} using TMA  get at
$T=1$ MeV larger clusters for $Y_{pg}=0.3$ and smaller  for $Y_{pg}=0.5$, and at
$T=5$ MeV larger clusters for $Y_{pg}=0.3$ and smaller  for $Y_{pg}=0.1$. It seems
that the proton fraction is not affecting the  size of the clusters in a linear
way, and this behavior is model dependent.  In fact, from Fig.\ref{fracyp} it is seen that
 the proton fraction with the largest fraction of nucleons in the clusters is
 $Y_{pg}=0.5$. However, the number of nucleons in the droplets also depends on
 their size. In Fig.\ref{radius2} we show how the size of the Wigner Seitz
 cell and droplet radius depend on the density and proton fraction. The number
of particles in the droplets is strongly dependent on the
these two radii.

In Fig. \ref{ratio2} we show for IU-FSU how the proton fraction in the droplets
depends on the temperature and on the global proton fraction. Temperature
makes droplets richer in protons and isospin asymmetry  reduces the proton
fraction in the droplets. In fact, temperature helps the evaporation of
neutrons from the droplets increasing the proton fraction in the clusters. On
the other hand, increasing the neutron fraction leads naturally to an increase
of neutrons in the clusters.
 These two effects were also obtained in the formalism
developed in \cite{raduta10} 
within a grand-canonical ensemble approach
(see left panels of Fig. 25 in Ref.~\cite{raduta10}).

\begin{figure}[ht]
  \centering
\begin{tabular}{c}
\includegraphics[width=0.9\linewidth]{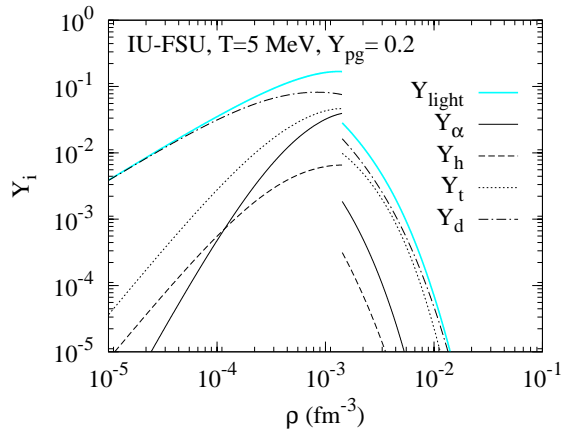}
\end{tabular}
\caption{(Color online) HPC case: the various contributions to $Y_{light}$ are shown explicitly for the IU-FSU parametrization at $T=5$ MeV, $Y_{pg}=0.2$.}
\label{lights}
\end{figure}

To complete the discussion, we now distinguish the contribution
of each type of light cluster, i.e., we decompose $Y_{light}$. In Fig.~\ref{lights} we show the fractions of nucleons belonging 
to $\alpha$'s, helions, tritons and deuterons for IU-FSU at $T=5$ MeV 
and $Y_{pg}=0.2$, as a choice that exemplifies the results we obtain.

Two typical sequences can be identified in the abundances of the various clusters: $d,t,h,\alpha$  at low baryonic density
and $d,t,\alpha,h$ at the higher total baryonic density at
which clusters are still present. Deviations from this behavior are found for lower 
temperature ($T=3.5$ MeV) or higher proton fraction ($Y_{pg}=0.5$).
In general, the dominant contribution comes from the deuterons. At very low density, the sequence reflects the ordering in size: the smaller clusters are more abundant.
The difference between $t$ and $h$, and more precisely the finding
that tritons are more abundant than helions, is due to two effects: 
globally there are more neutrons available and, moreover, the triton
is more bound. This last point explains why we find a difference among them also at $Y_{pg}=0.5$.

For a given temperature, the fraction of $\alpha$-particles at very low density increases with 
density much faster than the other clusters: this is due to their large
binding energy. As a consequence, the fraction of $\alpha$ particles may overcome
the fraction of other type of clusters at larger densities. At sufficiently  low temperatures, they are the most abundant
cluster in HC matter, even for the lowest densities.  In  HPC matter, their abundance is determined by the density, proton fraction,
 and temperature of the
background gas.

Concerning the dissolution density of each cluster, we find that
it increases with increasing temperature; on the other hand, the effect of a larger proton fraction is just to 
increase slightly the abundancy of the clusters (except for tritons), it does not alter their dissolution.

\begin{figure}[ht]
  \centering
\begin{tabular}{c}
\includegraphics[width=0.9\linewidth]{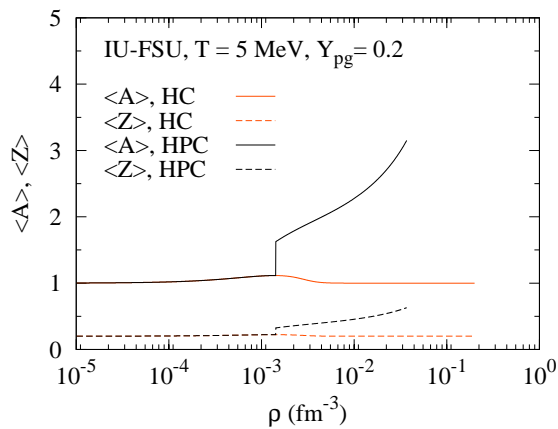}
\end{tabular}
\caption{(Color online) The average size and charge of the constituents present in HC and HPC for the IU-FSU parametrization at $T=5$ MeV, $Y_{pg}=0.2$.}
\label{averages}
\end{figure}

Finally, we show in Fig.~\ref{averages} the average number of nucleons and the average charge 
of the constituents in the HC and HPC cases. 
We consider as constituents the protons, neutrons, $\alpha$, helions, tritons, deuterons 
and the droplets. 
In the low density limit, the system is composed by free nucleons: in fact, $\langle A\rangle=1$ and $\langle Z\rangle=0.2$ in this limit. 
At intermediate densities, $\langle A\rangle$ increases slightly because
of the formation of light clusters. At even higher densities, the behavior is different
for the HC and HPC constructions. In the HC case, the clusters dissolve at
some density and we get again  $\langle A\rangle=1$, which means a gas of free nucleons. In the HPC case, droplets form, and the average
number of nucleons in a cluster increases very rapidly with the density.
Concerning the average charge, from the figure it is seen that for the droplets it also increases, but 
more slowly than $\langle A\rangle$: 
as already discussed, the isospin-asymmetry of the droplets increases with the density.

\section{Conclusions}

In the present paper we have investigated the effects caused by the
  explicit inclusion of four light clusters, namely, $\alpha$
  particles, deuterons, tritons and helions, in homogeneous and
  non-homogeneous nuclear matter at low densities.  This study is particularly
  important for the understanding of the composition of the inner
  crust of protoneutron stars.

We have chosen to calculate the above mentioned effects with four
different parametrizations of the NLWM because they are 
characterized by
different symmetry energy density dependencies, as seen from the
values of their slopes $L$ in Table II.

We have checked that the influence of the light clusters in the free
energy per particle is only noticeable at very small densities (up to 0.0025 fm$^{-3}$)
both in homogeneous and pasta phase matter. The inclusion of light
clusters lowers the free energy and their effect is  smaller in
the pasta phase range.

We have analyzed the fractions of nucleons at different temperatures
and different proton fractions. The results are model dependent, as
expected, and some of the differences are related to density dependence of the
symmetry energy.  In the following we identify some of the trends that were found: 
a) when temperature increases, the pasta phase appears at higher
densities, an effect already seen without the inclusion of the light
clusters;
b) the pasta phase low density gas, including light clusters, has a larger number of
particles at larger temperatures;
c)  {larger abundances of light clusters occur  at densities below
the pasta phase onset and their fractions are larger for more symmetric matter. 
Before the onset of the pasta phase their abundances decrease with
temperature, and the contrary occurs in the pasta phase;}
d) {within the coexisting-phases method adopted} in the present work, the onset
of the pasta phase occurs at lower densities for more asymmetric matter. We
believe this is due to the  zero thickness surface approximation and a lack of
self-consistency in the calculation of the surface properties. An
opposite trend was obtained  within a Thomas Fermi approach \cite{silvia10}.
 However, if we also consider the light clusters, the onset of
 clusterization starts at lower densities for the larger proton fractions. 

The number of nucleons and the number of charged particles inside the
pasta droplets were calculated. It was shown that:
a)  models with a smaller symmetry
energy slope have larger clusters, inside a larger Wigner-Seitz cell,  with a
larger number of particles and smaller proton fraction; 
b) the number of nucleons in the droplets has no linear
relation with the global isospin asymmetry of matter and density, showing a strong dependence on
the properties of the symmetry energy; 
c) the number of nucleons in the clusters decreases if temperature increases; 
d) the fraction of
protons in the clusters decreases with density, decreases for more asymmetric
matter and increases when  temperature increases.

We have shown that the composition and structure of the pasta phase is quite 
sensitive to the symmetry energy behavior at low densities.
It would be interesting to study the  transport properties and   neutrino opacity in the pasta phase
with light clusters, because these are  important quantities
in the cooling mechanism of protoneutron stars.

\section*{ACKNOWLEDGMENTS}
This work was partially supported by Capes / FCT n. 232 / 09 bilateral 
collaboration, by CNPq and FAPESC/1373/2010-0 (Brazil), by FCT and 
COMPETE (Portugal) under the projects PTDC/FIS/113292/2009 and CERN/FP/116366/2010 and  by Compstar, 
an ESF Research Networking Programme.
S. C. is supported by FCT under the project SFRH/BPD/64405/2009.

\section*{Appendix}

The surface tension is obtained by fitting 
the Thomas-Fermi results with the formula 
\begin{equation}
\sigma (x,T) = \tilde\sigma(x)\left[ 1-a(T)xT -b(T)T^2 -c(T)x^2 T\right]~, 
\end{equation}
where 
\begin{eqnarray}
\tilde\sigma(x)&=&\sigma_0\exp{(-\sigma_1 x^{3/2})}(1+a_1 x+ a_2 x^2\nonumber\\
&&+a_3 x^3 +a_4 x^4 + a_5 x^5 +a_6 x^6)\nonumber\\
a(T)&=& a_0 +a_1 T+ a_2 T^2 + a_3 T^3+ a_4 T^4 + a_5 T^5\nonumber\\
b(T)&=& a_0 +a_1 T+ a_2 T^2 + a_3 T^3+ a_4 T^4 + a_5 T^5\nonumber\\
c(T)&=& a_0 +a_1 T+ a_2 T^2 + a_3 T^3+ a_4 T^4 + a_5 T^5\nonumber\\
\end{eqnarray}

Clearly, $\tilde\sigma(x)$ is the surface tension at $T=0$, 
and $\sigma_0$ is its value at $T=0$ for symmetric matter.
Notice moreover that
there are a few more terms in these parametrizations as compared to those used 
in Ref.~\cite{pasta-alpha}.

\begin{table}[h]
\caption{Surface tension parameters fitted within the Thomas-Fermi 
approximation for NL3, NL3$\omega\rho$, FSU$_{\rm Gold}$ and 
IU-FSU parametrizations. The coefficients are for $T$ in MeV and $\sigma_0$ is
in MeV/fm$^{-2}$.}
\label{tabsig}
\begin{center}
\begin{tabular}{ccccc}
\hline
 NL3 & $\tilde\sigma(x)$ & $a(T)$ & $b(T)$ & $c(T)$  \\
\hline
  $\sigma_0$ & 1.12307 & - & - & - \\
  $\sigma_1$ & 20.7779 & - & - & - \\
  $a_0$ & - & 0.0121222 & 0.00792168 & - \\  
  $a_1$ & -5.84915 & 0.01664& -8.2504$\times 10^{-5}$  & - \\   
  $a_2$ & 138.839 & -0.00137266 & -4.59336$\times 10^{-6}$   & - \\   
  $a_3$ & -1631.42 & 4.0257$\times 10^{-5}$ & -2.81679$\times 10^{-7}$   & - \\     
  $a_4$ & 8900.34 & - & -  & - \\   
  $a_5$ & -21592.3 & - & -  & - \\   
  $a_6$ & 20858.6 & - & -   & - \\   
\hline
NL3$\omega\rho$ & $\tilde\sigma(x)$ & $a(T)$ & $b(T)$  & $c(T)$  \\
\hline
$\sigma_0$ & 1.12013   & - & - & - \\
$\sigma_1$ & 14.0774   & - & - & - \\
$a_0$      &   -       & -5.80451$\times 10^{-5}$  & 0.00725961              & -0.00259094 \\
$a_1$      & -2.15376  & 0.0233833                 & 0.000318409             &  -0.053756  \\
$a_2$      &  57.8455  & -0.00507732               &-0.000104941             &  0.0114598  \\
$a_3$      & -431.365  & 0.000490863               & 1.19645$\times 10^{-5}$ & -0.000354375 \\
$a_4$      & 1854.81   & -1.59473$\times 10^{-5}$  &-7.19099$\times 10^{-7}$ &-4.76451$\times 10^{-5}$ \\
$a_5$      & -3653.96  & -7.55062$\times 10^{-8}$  &1.62087 $\times 10^{-8}$ &2.28389$\times 10^{-6}$ \\
$a_6$      &  3214.82  & - & - & -  \\
\hline
FSU$_{\rm Gold}$ & $\tilde\sigma(x)$ & $a(T)$ & $b(T)$  & $c(T)$  \\
\hline
$\sigma_0$ & 1.1223   & - & - & - \\
$\sigma_1$ &-1.45717  & - & - & - \\
$a_0$      & -        & -0.0133789  & 0.00773356   & 0.0408077 \\
$a_1$      &-3.17729  & 0.0330912   & -0.000240406 & -0.0971609 \\
$a_2$      &-9.5121   & -0.00786564 & 4.52523$\times 10^{-5}$  & 0.0195288 \\
$a_3$      & 70.5609  & 0.000902286 & -7.64893$\times 10^{-6}$ & -0.00140166 \\
$a_4$      & -155.641 & -4.84828$\times 10^{-5}$ & 5.33346$\times 10^{-7}$ & 4.97386$\times 10^{-5}$ \\
$a_5$      & 154.691  & 9.56728$\times 10^{-7}$ & -1.45394$\times 10^{-8}$ & -1.20803$\times 10^{-6}$ \\
$a_6$      &-58.9476  & - & - & - \\
\hline
IU-FSU & $\tilde\sigma(x)$ & $a(T)$ & $b(T)$  & $c(T)$  \\
\hline
$\sigma_0$ & 1.16473   & - & - & - \\
$\sigma_1$ & -0.659167 & - & - & - \\
$a_0$      &   -       & 0.00404325                & 0.00767923               & 0.0066774  \\
$a_1$      & -2.25482  & 0.00828207                & -8.58068$\times 10^{-5}$ & -0.0514285 \\
$a_2$      & -5.64237  & -0.00153301               & 4.43918 $\times 10^{-7}$ & 0.00949505 \\
$a_3$      & 37.8471   & 7.26763$\times 10^{-5}$   & -5.44453$\times 10^{-7}$ &-0.000427613 \\
$a_4$      &-81.6617   & - & - & -  \\
$a_5$      & 81.2696   & - & - & -  \\
$a_6$      &-31.0227   & - & - & -  \\
\hline
\end{tabular}
\end{center}
\end{table}

\end{document}